# Strengthening from dislocation restructuring and local climb at platelet linear complexions in Al-Cu alloys

Pulkit Garg [a], Daniel S. Gianola [b], Timothy J. Rupert [a,c,*]

[a] Department of Materials Science and Engineering, University of California, Irvine, CA 92697, USA
[b] Materials Department, University of California, Santa Barbara, CA 93106, USA
[c] Department of Mechanical and Aerospace Engineering, University of California, Irvine, CA 92697, USA
* To whom correspondence should be addressed: trupert@uci.edu

**Abstract**

Stress-driven segregation at dislocations can lead to structural transitions between different linear complexion states. In this work, we examine how platelet array linear complexions affect dislocation motion and quantify the associated strengthening effect in Al-Cu alloys using atomistic simulations. The presence of platelet complexions leads to the faceting of the dislocations, with nanoscale segments climbing upwards along the platelet growth direction, resulting in a complex configuration that restricts subsequent dislocation motion. Upon deformation, the leading partial dislocation must climb down from the platelet complexions first, followed by a similar sequence at the trailing partial dislocation, in order to overcome the precipitates and commence plastic slip. The dislocation depinning mechanism of linear complexions is strikingly different from traditional precipitation-strengthened alloys, where dislocations overcome obstacles by either shearing through or looping around obstacles. The critical shear stress required to unpin dislocations from platelet complexions is found to be inversely proportional to precipitate spacing, which includes not just the open space (as observed in Orowan bowing) but also the region along the platelet particle where climb occurs. Thus, linear complexions provide a new way to modify dislocation structure directly and improve the mechanical properties of metal alloys.

**Keywords:** Dislocations, Segregation, Precipitation, Strengthening

**Introduction**

Dislocations are linear defects that play a pivotal role in determining the mechanical properties of metals as the direct carriers of plasticity. As they move through the crystal lattice, dislocations interact with one another and various microstructural features such as solutes, precipitates, and grain boundaries, determining the resultant strength of metallic materials [Kuhlmann-Wilsdorf 1989, Soboyejo 2002]. The addition of alloying elements to introduce second phase precipitates as obstacles to dislocation motion is one of the most commonly implemented methods to improve the strength and hardness of structural alloys [Antion et al. 2003, Ardell et al. 1976, Gladman 1999, Kohler et al. 2004, Nie 2012]. The presence of supersaturated solid solutions in the grain interior can lead to the homogeneous nucleation of precipitates that significantly contribute to the strengthening of structural alloys [Liu et al. 2017, Robson et al. 2003]. On the other hand, local variations in structure and stress state at crystalline defects can also attract alloying elements and drive the heterogeneous nucleation of precipitates [Feng et al. 2010, Medouni et al. 2021, Miller 2006]. The mechanical behavior of precipitation-strengthened alloys is dependent on a variety of parameters such as the precipitate morphology (shape, size, orientation, etc.), the nature of the precipitate/matrix interface, and the details associated with the dislocation-precipitate interactions [Adlakha et al. 2019, Ahmadi et al. 2014, Gerold and Haberkorn 1966].

In general, dislocations are observed to overcome second phase precipitates by a shearing mechanism if the precipitates are small and coherent with the matrix material, while larger precipitates become incoherent and must be overcome by the Orowan looping mechanism [Ardell 1985, Hirsch 1957, Orowan 1948]. Guinier-Preston (GP) zones are often the precursors to precipitation that first form during the aging of Al-Cu alloys [Gerold 1988, Starink and -M

ZAHRA 1997], hindering dislocation motion and thus increasing the macroscopic strength of the material [Hornbogen 2001]. The details of dislocation-precipitate interactions are typically determined by the misfit stresses and the difference in elastic modulus between the precipitate and the matrix [Takahashi and Ghoniem 2008]. For example, Singh and Warner [Singh and Warner 2010] observed a variety of mechanisms such as dislocation looping, shearing, cross-slip, and defect nucleation upon dislocation–GP zone interaction, finding significant variations in the shear stress required to overcome the precipitates that depended on the GP zone size, orientation, and offset from the dislocation glide plane. The stress required for an edge dislocation to move past a GP zone in Al-Cu also depends on the dislocation-precipitate orientation such that the maximum stress is required when the GP zones lie at an angle of 60 degrees with the slip plane [Singh and Warner 2013, Singh et al. 2011]. Thus, these microstructural features significantly contribute to the strength and plasticity of Al alloys and demonstrate a rich variety of dislocation-obstacle interaction physics.

Linear complexions present an even more direct method to manipulate dislocations by altering their local structure and chemistry, opening up a new pathway to engineer defects and modify the mechanical behavior of metal alloys. Linear complexions are thermodynamically-stable, nanoscale phases that form upon heterogenous nucleation of precipitates at dislocations where the particles remain constricted to the high-stress regions near the defect. These features are structurally and/or chemically different from the matrix and were first reported by Kuzmina et al. [Kuzmina et al. 2015] in Fe-Mn, where local regions with face-centered cubic (FCC) structure and an austenite composition were observed only near the dislocations in the body-centered cubic matrix. The formation of linear complexions has also been reported variety of material systems such as irradiated low alloy reactor pressure vessel steels [Odette et al. 2021], sulfide ores [Fougerouse et al. 2021], and Pt-Au alloys [Zhou et al. 2021]. Linear complexions have been observed to lead to static strain aging in Fe-Mn steels [Kwiatkowski da Silva et al. 2017] and also have been predicted to impart higher resistance to dislocation motion as compared to solid solution strengthening in FCC alloys [Singh et al. 2023]. The presence of linear complexions can alter the dislocation strain fields and has been reported to influence the morphology of spinodal precipitates at dislocations even with the global composition outside the spinodal limit [Varma R. et al. 2023]. Categorized based on their effect on the dislocation core, three different types of linear complexions have been predicted to exist in FCC alloys [Turlo and Rupert 2020], with platelet array linear complexions restructuring the dislocation core due to their growth crystallography. The Al-Cu alloy system is predicted to contain such platelet linear complexions, with the particles being disk-shaped precipitates containing Cu atoms that grow along the dislocation length and out of the slip plane. The local structure of platelet complexions is very similar to that of GP zones, which have been extensively observed during the aging of Al–Cu bulk alloys [Gornostyrev and Katsnelson 2015, Liu et al. 2017, Ringer and Hono 2000], making these linear complexions likely to restrict dislocation motion and improve alloy strength as well. The mechanisms by which dislocations overcome GP zones and the associated strengthening of Al-Cu alloys are very well understood, while the similar impact of platelet linear complexions has not been investigated to date.

In this work, we study the deformation mechanisms and strength scaling laws associated with platelet array linear complexions in Al-Cu alloys using atomistic simulations. Al-Cu models with linear complexions were first created, to isolate platelet-shaped precipitates at a single pair of Shockley partial dislocations. The platelet complexions are localized to the compressed regions near the dislocation, relaxing these stresses and driving the reorganization of the dislocation structure. The platelet complexions grow out of the slip plane, pushing the dislocation to vary its line direction and causing faceting of the dislocations. The shifted dislocation segments primarily have edge character and locally climb along the growth direction of platelet complexions, creating a complex configuration that restricts dislocation motion. The mechanism by which dislocations overcome the platelet complexions is strikingly different from classical precipitation-strengthened alloys, as the dislocations must climb downward and breakaway from the platelet complexions for plasticity to commence. The critical shear stress required for dislocation breakaway is significantly affected by the size and spacing of the platelet complexions and does not follow the classical Orowan strengthening equation. Overall, this work demonstrates that linear complexions provide a new pathway for the development of high performance Al alloys with improved mechanical properties.

**Computational Methods**

Linear complexion models were created with hybrid Monte Carlo (MC)/molecular dynamics (MD) simulations using the Large-scale Atomic/Molecular Massively Parallel Simulator (LAMMPS) code [Thompson et al. 2022], with a 1 fs integration time step for all MD simulation components. An embedded atom method potential for the Al-Cu system [Cheng et al. 2009] was used in this study that can reproduce important features of the bulk phase diagram, specifically the existence of the GP zones, other metastable intermetallic phases, and the stable secondary phases. This interatomic potential has also been successfully used to investigate linear complexions and grain boundary complexions in previous studies [Hu and Rupert 2019, Singh et al. 2023, Turlo and Rupert 2020]. We note that this interatomic potential underestimates the stacking fault energy of this material, leading to larger partial dislocation spacings than experimental observations.

Atomic configurations were analyzed and visualized using the common neighbor analysis (CNA) [Faken and Jónsson 1994] and dislocation analysis (DXA) [Stukowski et al. 2012] methods within the open-source visualization tool OVITO [Stukowski 2009].

First, a pure Al sample with two edge dislocations of opposite character was created, where each of the dislocations relaxed into Shockley partial dislocations with a stacking fault between them upon equilibration at 250 K using a Nose-Hoover thermo/barostat at zero pressure. The pure sample was then doped with 0.3 at.% Cu at 250 K using the hybrid MC/MD method, leading to the nucleation of linear complexions in the form of platelet particles that grow out of the slip plane (additional details are provided in Ref. [Turlo and Rupert 2020]). The MC steps performed atom swaps in a variance-constrained semi-grand canonical ensemble [Sadigh et al. 2012] with MC switches occurring after every 100 MD steps, with the global composition fixed to 0.3 at.% Cu by adjusting the chemical potential difference during the simulation. This typ of hybrid MC/ MC approach has been successfully used to examine the role of grain boundary character on solute segregation behavior and study its impact on the enhanced thermal stability of nanocrystalline Pt-Au alloys [Barr et al. 2021], as well as to determine the impact of local chemical ordering on dislocation activity [Chen et al. 2023] and study the enhanced radiation damage tolerance of high entropy alloys [Zhang et al. 2023]. The hybrid MD/MC simulations were implemented here as they allow a realistic configuration of platelet complexions to be created through chemical and structural relaxation of the system. The resulting platelet linear complexions form due to the local stress field of the dislocation, without any assumptions or preconceived notions about what the platelets may look like. To determine the linear complexion strengthening effect and associated deformation mechanisms, one of the partial dislocation pairs was isolated and all except one complexion precipitate per partial dislocation line were dissolved in order to obtain the model system shown in Fig. 1. The dissolution was accomplished by changing atom type from Cu to Al, followed by the equilibration of the isolated linear complexion sample at 250 K using a micro-canonical ensemble (NVE) for 100 ps. Fig. 1(a) shows an isometric view of the equilibrated linear complexion sample with the X-axis oriented along the [110] direction (Burgers vector of a full dislocation), the Y-axis oriented along [1$\bar{1}$1] direction (slip plane normal), and the Z-axis oriented along the [$\bar{1}$12] direction (dislocation line length). The simulation cell is oriented along the close-packed slip system in FCC metals and the implemented geometry is similar to the simulation cells used in the literature [Adlakha et al. 2019, Singh and Warner 2010, Singh and Warner 2013] to simulate classical dislocation-obstacle interactions and their associated strengthening mechanisms. The FCC Al atoms have been removed for clarity in all images presented in this paper and the platelet complexions on the left and right partial dislocation are denoted as *A* and *B*, respectively. The simulation cell shown in Fig. 1 is approximately 60 nm long (X-direction), 15 nm tall (Y-direction), and 10 nm thick (Z-direction), containing ~500,000 atoms, with non-periodic shrink-wrapped boundary condition prescribed along the Y-direction and periodic boundary conditions along the X and Z directions. The thickness of the simulation cell in the Z-direction defines the platelet spacing (*L*), the distance between the centers of two identical platelet complexions along the dislocation line length. The L value is systematically increased from 10-40 nm by replicating the simulation cell along the periodic Z-direction (dislocation line length). In the larger simulation cells, only the original complexion precipitate pair was retained while all other complexion precipitates created by replication were dissolved by changing the atom type from Cu to Al, followed by the equilibration of the samples at 250 K for 100 ps using a micro-canonical ensemble (NVE). Fig. 1(b) and (c) show the same sample where *L* = 10 nm from different views along the dislocation line length and slip plane normal direction, respectively. Both the platelet complexions grow out of the slip plane at the partial dislocations and consist of a single atomic layer of Cu atoms at each of the partial dislocations while maintaining their platelet/disk shape after equilibration, consistent with previous structural simulations [Turlo and Rupert 2020].

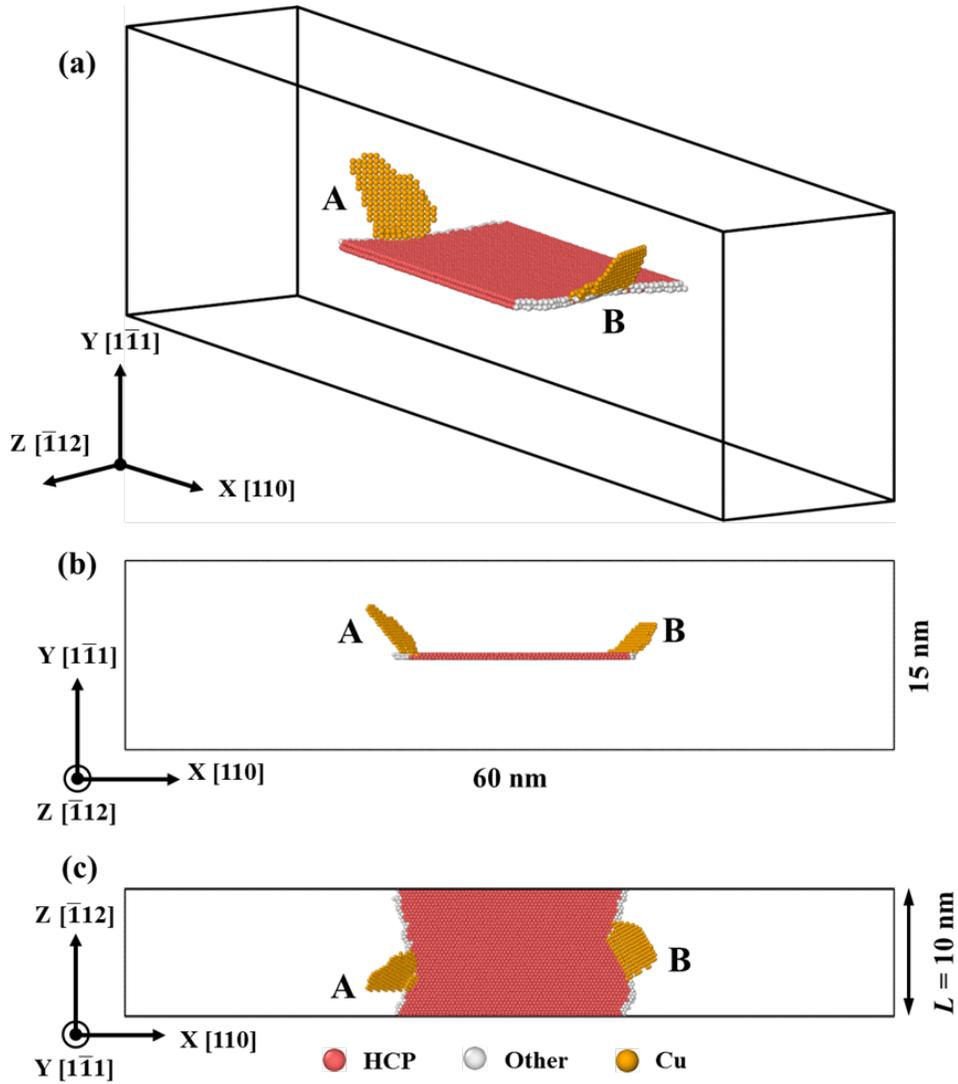

Fig. 1. (a) Isometric view of the Al-Cu linear complexion sample with $L$ = 10 nm where platelet complexions (A and B) have nucleated at each of the partial dislocations, which are separated by a stacking fault. Views along (b) the dislocation line length and (c) the slip plane normal direction show the platelet complexions growing out of the slip plane at each of the partial dislocations. Atoms are colored according to their local atomic structure (HCP, other, or Cu), while all of the Al atoms with FCC crystal structure have been removed for clarity.

The critical shear stress ($\tau_{crit.}$) required to move the partial dislocations away from the platelet linear complexions was investigated to understand mechanical properties. The top and bottom regions (~0.5 nm) along the Y-direction were fixed and a constant shear stress was applied to the Y-axis face, along the X-direction. The samples were held under the applied shear stress for 100 ps while the dislocation positions were recorded to determine if dislocation motion occurred. The applied shear stress was systematically increased, starting with increments of 50 MPa, followed by smaller changes of 5 MPa, to obtain $\tau_{crit.}$ corresponding to the minimum shear stress that resulted in the depinning of the dislocations from the platelet complexions and moved through the cells. Since the leading and trailing partial dislocations can move at different stress levels (and indeed they always did), both critical events were isolated for each simulation variation. Such stress controlled simulations have been commonly used in literature to measure the critical shear stress for dislocation motion and/or dislocation mobility as a function of applied stress as they provide sufficient time to reach the equilibrium state and do not introduce any new defects artificially [Krasnikov et al. 2020, Mishra et al. 2021, Yin et al. 2021].

**Results and discussion**

First, we examine the platelet linear complexions themselves to understand their influence on the local stresses and structure of the dislocations. Fig. 2(a) shows the distribution of local hydrostatic stress in the pure Al sample with a pair of Shockley partial dislocations. The partial dislocations cause local stress concentrations with tensile stresses in the direction where the extra half-plane was removed (i.e., below the slip plane here) and compressive stresses occur in the other region (i.e., above the slip plane here), consistent with the previous studies [Soleymani et al. 2014].

Upon doping of the pure Al sample, the Cu atoms segregate to the compressive side of the partial dislocations due to their low atomic volume as compared to Al and form the platelet linear complexions, as shown in Fig. 2(b). The formation of platelet linear complexions relaxes the compressive hydrostatic stresses, altering the stress field at the partial dislocations.

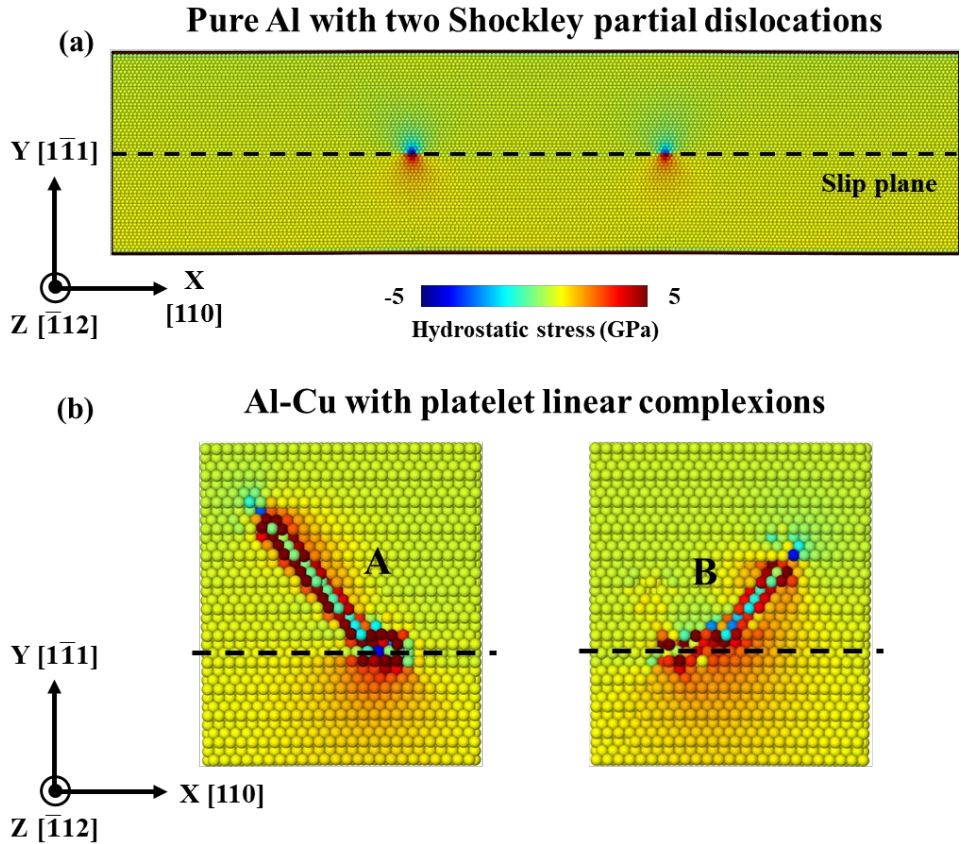

**Fig. 2. Distribution of the local hydrostatic stress in (a) the pure Al sample containing Shockley partial dislocations and (b) at the platelet linear complexions in the Al-Cu sample. Segregation and complexion nucleation occurs at the compressed region above the slip plane.**

The growth of platelet complexions out of the slip plane and in tilted directions necessitates the identification of the orientation relationships between the dislocations and the platelet particles. Fig. 3(a) shows the platelet linear complexion sample as viewed along the slip plane normal (Y-direction). The platelet complexions start growing at the partial dislocations and experience directional growth into another plane inclined at an angle from the original dislocation line. Both the platelets are oriented at an angle of 30º from the original dislocation line (view along the slip plane normal in Fig. 3(b)) and at an angle of 54.5º with the slip plane (view along the local dislocation line, $\vec{t}$, in Fig. 3(c)), such that they appear as mirror images of one another. The growth crystallography of the platelet linear complexions is similar to that observed for GP zones in Al-Cu binary alloys, where the soft elastic response of Cu along the energetically favored direction promotes monolayer clustering of Cu atoms into plate-like regions [Stroev et al. 2018, Wolverton 1999]. The orientation relationship between the dislocation slip plane and platelet complexions is also consistent with the dislocation-obstacle orientation that provided the maximum strengthening from GP zones [Singh and Warner 2013]. However, the platelet complexions differ from traditional GP zones in a number of important ways. For example, the platelet complexions are also inclined with respect to the dislocation line length, a feature not observed for GP zones. In addition, platelet linear complexions can be thicker in the out-of-plane direction near the dislocation line where they originate, as shown in Fig. 5.

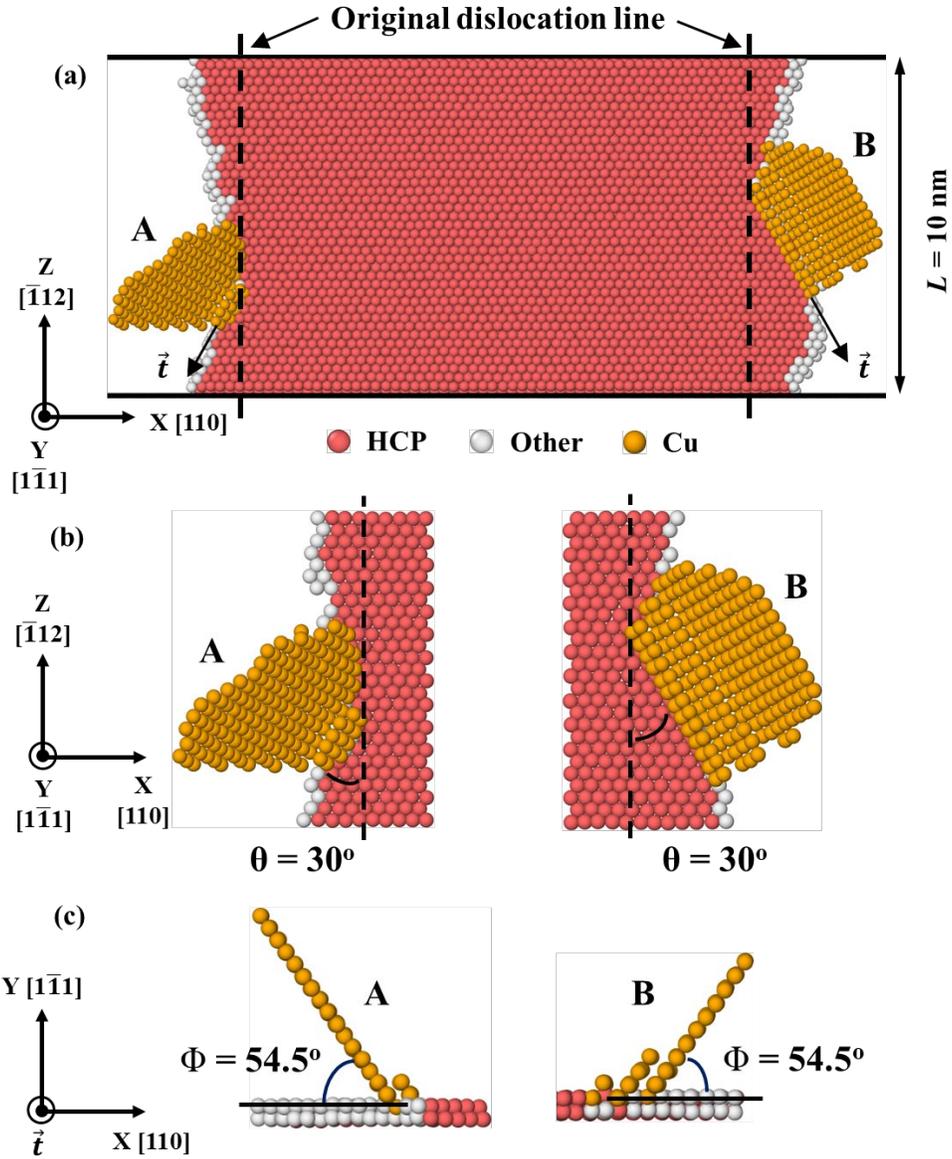

Fig. 3. (a) View of the platelet array linear complexion sample with $L$ = 10nm, viewed normal to the slip plane where the original Shockley partial dislocation line directions are marked by dashed black lines. The orientation relationship of the platelet particles with respect to (b) the original dislocation line directions (Z-direction) and (c) the slip plane (X-direction). Both the platelet complexion precipitates have an identical orientation relationship with other defect structures in this system.

The growth of two-dimensional platelets at the partial dislocations alters the original dislocation lines and leads to faceting, as shown in Fig 4. Most obvious in Fig. 4(a) are the two relatively straight facet lengths where the partial dislocations intersect with the platelets, which is closely related to the overall size of each platelet particle and should be relevant to any measured strengthening effects, as this distance marks the region of defect-obstacle interaction. This interaction region between the dislocations and obstacles plays a critical role in determining the mechanism by which dislocations overcome traditional precipitate obstacles [Adlakha et al. 2019, Nie and Muddle 1998]. The facet length of platelet particle A ($h_A$) is 1.1 nm and that of the platelet particle B ($h_B$) is 3.9 nm. Fig. 4(b) shows the platelet particles where the partial dislocation lines are colored according to their dislocation character. Examination of the dislocation character shows that the locally faceted dislocation segments shift out of their glide planes along the platelet complexions and primarily have edge character. The dislocations remain in the original slip plane otherwise. In contrast, the dislocation segments away from the platelet particles have neither full edge nor full screw dislocation character. This change in the dislocation character is caused by the platelet complexion stress field that distorts the dislocation core and changes the way the dislocation interacts with the surrounding atoms. Such reorientation of edge dislocations to dislocations with mixed dislocation character in the vicinity of precipitates is consistent with other recent observations in bulk FCC alloys [Kioussis and Ghoniem 2010, Xu et al. 2019].

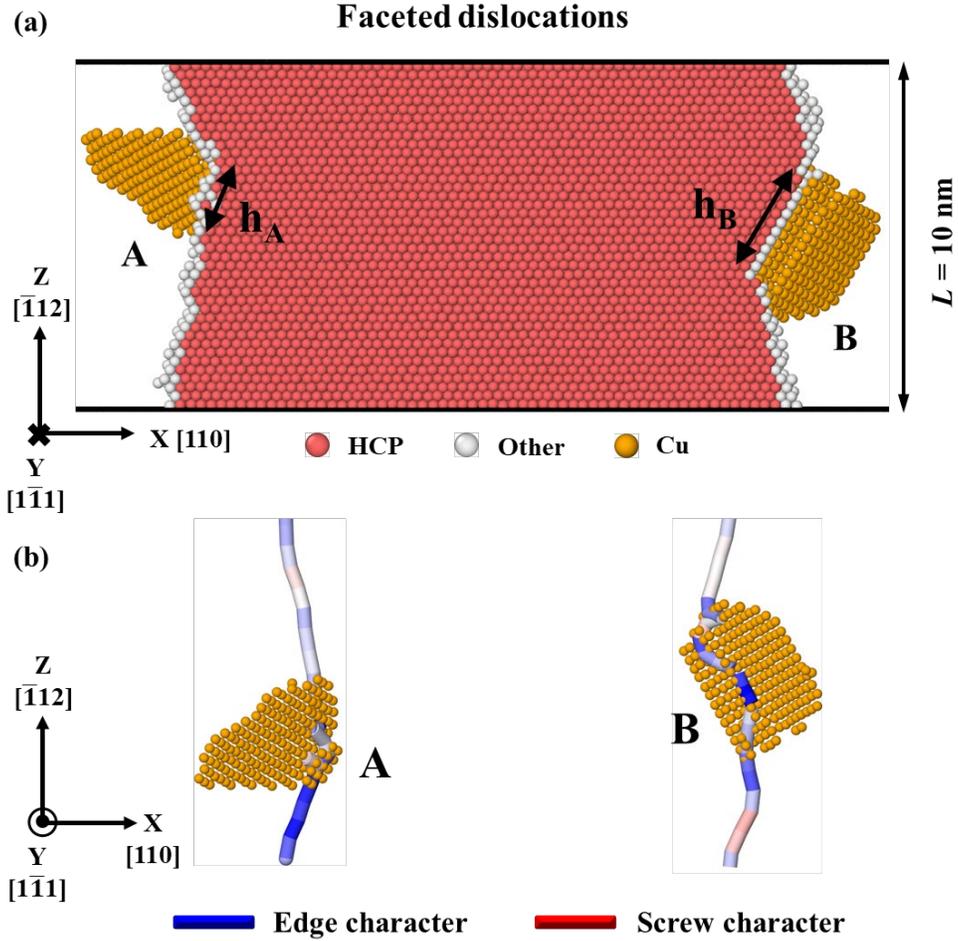

**Fig. 4.** (a) The platelet linear complexion sample with $L = 10$ nm, viewed along the slip plane normal where faceting of the dislocation line is obvious. The distance along which the dislocation facet interacts with the platelet can serve as a measure of its size, $h$. (b) Analysis of the dislocation shows that the faceted dislocation segments with edge character shifted out of their glide planes along the platelets.

The observation of faceted dislocation segments with edge character and the possibility that they are shifting out of their slip plane along platelet complexions highlights the importance of local dislocation structure. Fig. 5 presents edge-on views of the platelet particles and the nearby partial dislocations, with the Cu atoms in the platelets colored according to their distance above the slip plane to highlight the atomic layers where dislocation segments have climbed. The platelet complexions form coherent interfaces with the matrix along the (100) plane [Turlo and Rupert 2020], which is a non-dominant slip system in FCC crystals, thereby restricting dislocation cross-slip and promoting dislocation climb along the precipitates. Finally, the relative line direction of the dislocation segments identified as screw and edge are orthogonal, as expected from fundamental dislocation theory, providing additional confirmation of a dislocation climb mechanism. At both particles, the Shockley partials have moved a few atomic layers upward along the platelet growth direction. The upward motion is more pronounced for platelet B, which is the bigger of the two particles in this system. In this figure, the dislocation line is again colored according to the dislocation character. The shifted dislocations predominantly have an edge character although some screw character remains, indicating a mixture of cross-slip and climb. We primarily focus on the climb component since this should be (1) the dominant behavior due to the majority edge character [Baker and Curtin 2016, Thomson and Balluffi 1962] and (2) the climb is generally a more rate-sensitive mechanism [Vevecka-Priftaj et al. 2008]. A similar local climb mechanism has been observed in traditional precipitation-strengthened alloys, where dislocation segments were observed to shift onto the precipitate along the precipitate-matrix interface [Brown and Ham 1971, Clark et al. 2005, Krug and Dunand 2011]. In general, the platelet array linear complexions create a complex, non-planar configuration out of a once simple partial dislocation pair.

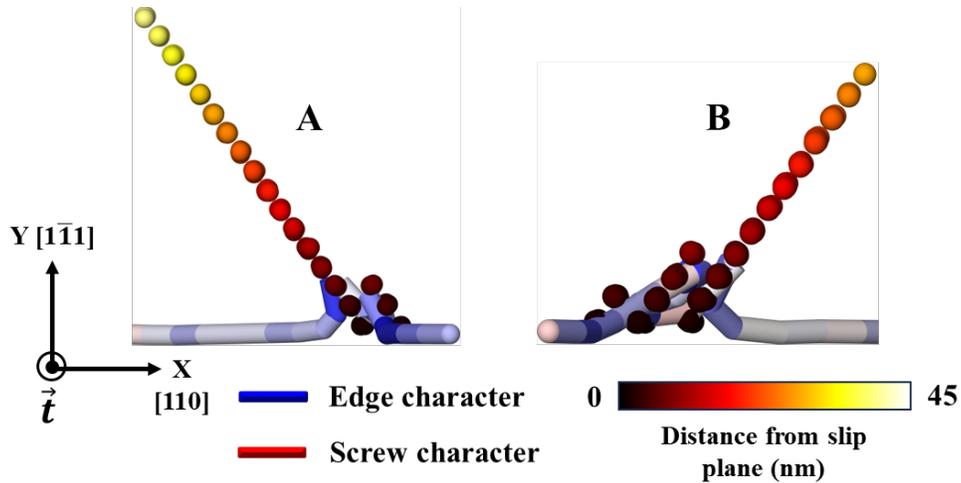

Fig. 5. Edge-on views of the platelet linear complexions with $L$ = 20 nm showing the dislocation segments that have shifted out of their glide plane and primarily climbed along the platelet complexions. All the Al atoms are removed for clarity and Cu atoms are colored according to their position in the Y-direction. The dislocation lines are colored according to dislocation character, showing that a majority edge configuration, although some minor screw character persists.

Next, it is pertinent to understand how the dislocation moves away from this configuration. Fig. 6 shows the process of the partial dislocation climbing down and subsequently breaking away from platelet B during the deformation of the Al-Cu complexion sample with $L$ = 20 nm at the critical shear stress. After loading is applied, the partial dislocation segment that had climbed begins to move downward, one atomic layer at a time. After 4 ps, the dislocation segment has climbed down by one atomic layer (view along the local dislocation line) while the dislocation line along the slip plane begins to bow around the platelet complexion (view along the slip plane normal), as shown in Fig 6(a). The dislocation portion along the platelet complexion continues to climb down the atomic layers while the dislocation segment along the slip plane continues to bow as time progresses (Figs. 6(b) and (c)). Eventually, the partial dislocation returns to the original slip plane and is able to begin breaking away from the platelet complexion in Fig. 6(d). Figs. 6(e) and (f) show where the segment that was originally stuck is finally all moving and when it has begun to straighten out, respectively.  An identical mechanism is observed for the partial dislocations that climb down and break away from platelet A.  Although the dislocation bowing shape and mechanism look somewhat similar to what occurs at traditional precipitates, the need for this layer-by-layer climb is unique to these platelet array linear complexions.  Additionally, the traditional mechanisms do not account for dislocation interaction with the precipitate interior and require a high density of precipitates oriented along the maximum strengthening direction to optimize dislocation-precipitate interactions.  However, such limitations do not exist for platelet array linear complexions that are equilibrium states localized to the dislocations, naturally oriented along the maximum strengthening direction, and maintain their shape and structure after dislocation breakaway.  This is also strikingly different from dislocations overcoming the precipitates by climbing into the bulk regions without interaction with the precipitates [Singh and Warner 2010, Xiang and Srolovitz 2006], as the dislocations climb along the precipitates in platelet array linear complexions and continuously interact with them while climbing down and breaking away.

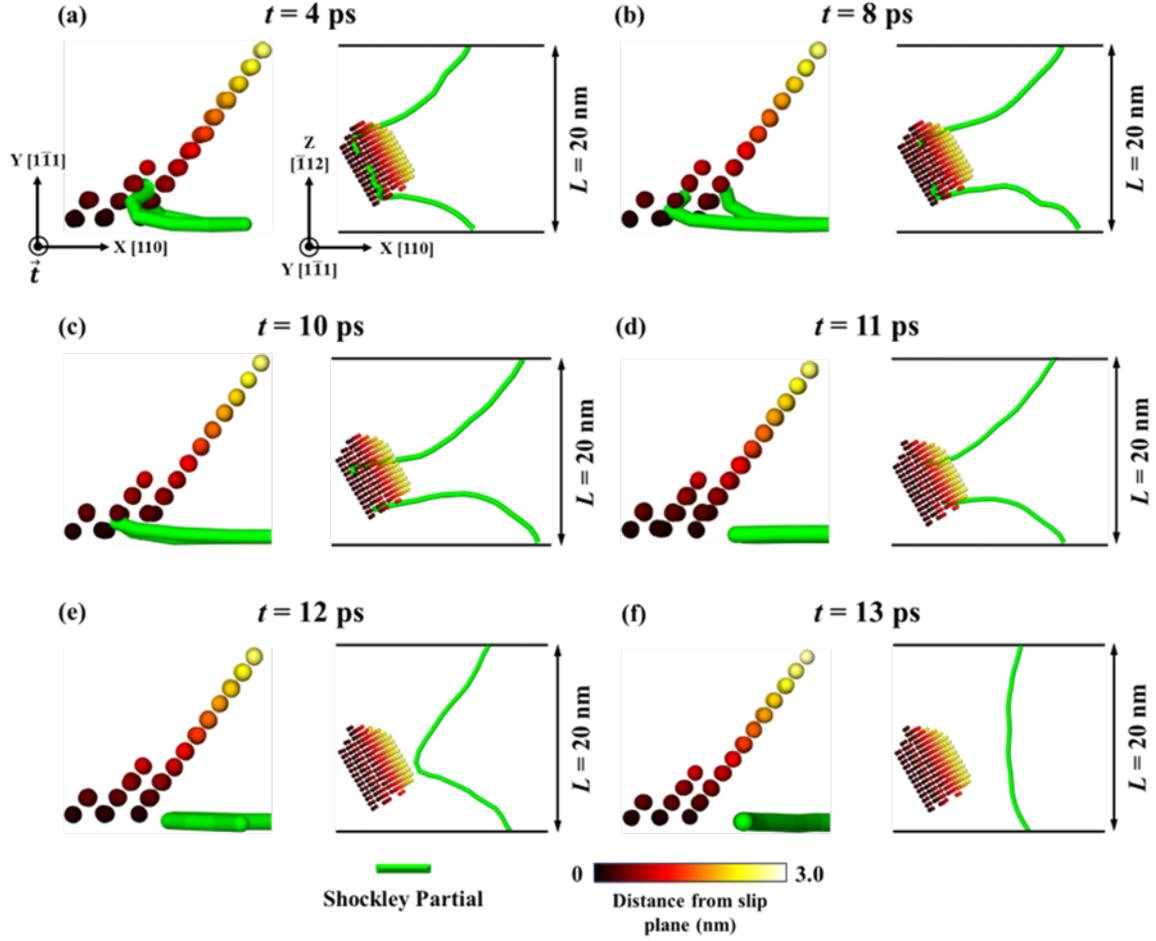

**Fig. 6.** Analysis of the partial dislocation climbing down and then breaking away from platelet B in the Al-Cu complexion sample with *L* = 20 nm when the critical shear stress has been applied. All of the Al atoms have been removed for clarity and Cu atoms are colored according to their position in the Y-direction.

Next, the critical shear stress values required for the partial dislocations to overcome the platelet linear complexions are extracted, in order to develop a strength scaling law. In the model system used in this study, the two platelet particles have very different sizes, which we hypothesize to strongly impact strength and which are described in terms of the facet length shown in Fig. 4. The symmetry of the simulation cell allows the shear stress to be applied in such a way that it drives dislocation motion in either the positive or negative X-direction, referred to as *forward shear loading* ($\tau_F$) and *reverse shear loading* ($\tau_R$), respectively. The different loading configurations, as well as the different critical plastic events, are shown in Fig. 7. Figs. 7(a) and (b) show the leading partial dislocations breaking away from platelet B with $h_B = 3.9$ nm during forward shear loading and platelet A with $h_A = 1.1$ nm during reverse shear loading, respectively. In all cases, the leading partials move away first while the trailing partials remain pinned. In fact, higher applied shear stress levels are needed in all cases to drive the trailing partial way from the platelets, as compared to the leading partials. While both events are interesting and will be discussed, it is important to remember that the partial pair would remain stuck at the obstacle pair until the trailing partial can break away, making the movement of the trailing partial event most relevant to plastic deformation. It is only at this point that the partial pair has fully released from the obstacle. Representative examples of motion of the trailing partials from the platelets are shown in Figs. 7(c) and (d). While the dislocation line length was varied, an identical dislocation breakaway mechanism was observed for all iterations.

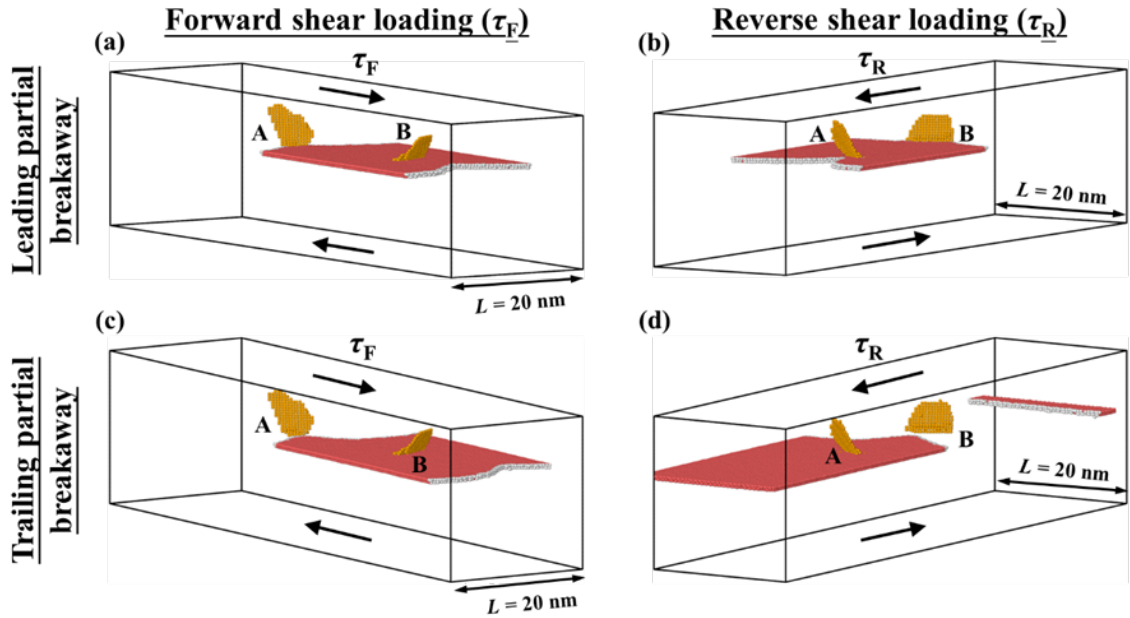

**Fig. 7.** Critical dislocation breakaway events in an $L = 20$ nm sample loaded in both forward (left column) and reverse shear loading (right column). The leading partial dislocations are always the first to overcome the platelet obstacles, with (a) platelet B overcome first during forward shear loading and (b) platelet A overcome first during reverse shear loading. The trailing partial dislocations always follow and require higher applied shear stresses, with (c) platelet A overcome last during forward shear loading and (d) platelet B overcome last during reverse shear loading. All of the FCC atoms have been removed for clarity.

Finally, Fig. 8 compiles all of the measured values for $\tau_{crit.}$ for the breakaway of the leading and trailing partial dislocations during both the forward and reverse shear loading conditions, with the data shown as a function of $L$. This figure clearly shows that there are two distinct populations of values for either leading or partial dislocation motion. That is, one direction is significantly stronger than the other for a given mechanism. For example, for leading partial dislocation breakaway, the critical shear stresses are higher for the forward loading conditions (Fig. 8(a)). Referencing the configurations shown in Fig. 7, one can see that this is the case because the partial dislocation has to overcome the larger platelet B when loaded in this direction.

Following the same logic, the motion of the trailing partial is more difficult during reverse shear loading because, again, the critical event is overcoming the obstacle that is platelet B. The increased difficulty of climbing downward for a larger distance therefore manifests as a noticeably stronger obstacle. The data in Fig. 8 can be described using a modified strengthening law:

$$\tau_{crit.} = \beta_{size} \cdot \frac{Gb}{L} \tag{1}$$

where $\beta_{size}$ is a pre-factor associated with the platelet size, $G$ is the shear modulus of the crystal through which the dislocation moves, $b$ is the Burgers vector, and $L$ is the platelet spacing. Equation 1 resembles the Orowan equation [Bacon et al. 1973] but varies in a number of important ways. For example, the classical Orowan equation scales with $(L - 2r)$, where $r$ is the radius of a spherical particle. This means that only the open space between obstacles matters since this is the region through which bowing can occur, while the actual size of the particle plays no role. For the platelet array linear complexions, the distance along the obstacle contributes as well, since this length also bows (see Fig. 6(e)). Equation 1 also has a pre-factor which directly accounts for the particle size and its effect on the dislocation moving out of the original slip plane. More movement and a stronger strengthening effect occur for the larger platelet, suggesting that this pre-factor should increase as platelet size increases. Figs. 8(a) and (b) show that Equation 1 fits the data well, with the best fit found for the more important dataset in Fig. 8(b) associated with the critical event for plastic flow. Matching physical intuition, $\beta_{size}$ is indeed larger whenever it is associated with a partial breaking away from the larger platelet B.

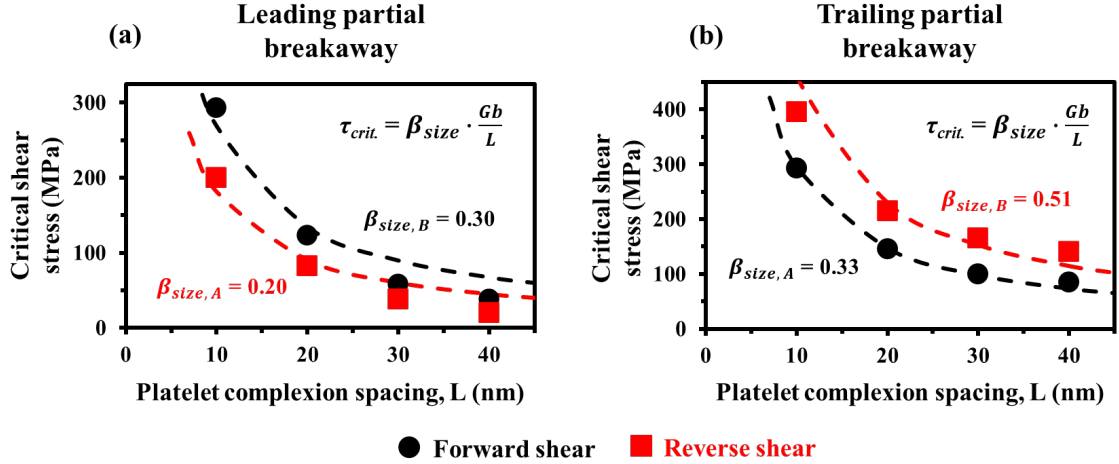

Fig. 8. The critical shear stress ($\tau_{crit.}$) required for (a) leading and (b) trailing partial dislocation breakaway from platelet array linear complexions, presented as a function of platelet spacing.

**Summary and Conclusions**

In this work, the influence of platelet array linear complexions on dislocation motion was examined to understand critical deformation mechanisms and the strengthening effect of these features. The following conclusions can be drawn from this study:

- In Al-Cu, platelet linear complexions are localized to the compressive stress regions near the dislocations, reducing this stress field as the new structure nucleates.

- Platelet linear complexions grow out of the slip plane and can cause faceting of the dislocation lines. The faceted segments along the obstacles primarily have edge character, whereas the dislocation segments away from the platelet complexions have more evenly mixed edge and screw dislocation character.

- The Shockley partial dislocations remain in the slip plane along most of the line length, except where they shift out of their glide plane along the platelet particles. These segments climb a few atomic layers along the platelet complexions, resulting in a complex configuration that is able to efficiently restrict subsequent dislocation motion.

- The partial dislocations must climb downwards to reach the original slip plane before they can move past the platelet obstacles. Thus, the mechanism by which dislocations overcome platelet linear complexions is strikingly different from traditional precipitation-strengthened alloys where dislocations overcome the obstacles by shearing, looping, or a combination of the two mechanisms.

- The critical shear stress required for dislocation breakaway is inversely proportional to platelet complexion spacing and strongly affected by the size of the platelet particles. Rather than only considering the open gaps as the Orowan equation does, the total distance along the dislocation line must be considered as the facet touching the platelet participates in plasticity.

Overall, this work provides a new understanding of the effect of platelet array linear complexions and their effect on the mechanical behavior of metal alloys. The dislocation restructuring due to the nucleation of platelet complexions and the observed dislocation climb along the precipitates provide new obstacles to dislocation motion, thus strengthening the material. The presence of dislocation climb along the platelet complexions may also increase the temperature dependence, strain rate sensitivity, and strain hardening behavior of these alloys, as dislocation climb is an extremely rate sensitive deformation mechanism and movement out of the original glide plane can increase dislocation-dislocation interactions. Platelet linear complexions should be an additional tool for increasing the strength of Al alloys under high strain rate conditions, as well as for manipulating strain hardening rates and the evolution of dislocation density during plastic deformation. In the future, it will also be important to seek experimental verification of the mechanisms by which dislocations overcome the platelet complexions and also analyze the impact of precipitate morphology on strengthening derived from platelet linear complexions. These linear complexions can provide efficient strengthening for Al-Cu alloys and offer future opportunities for defect engineering in light metal alloys, opening up even broader possibilities.

**Availability of data and materials**

The data that support the findings of this study are available within the article.

**Competing interests**

The authors declare that they have no known competing financial interests or personal relationships that could have appeared to influence the work reported in this paper.

**Funding**

This research was sponsored by the Army Research Office under Grant Number W911NF-21-1-0288. The views and conclusions contained in this document are those of the authors and should not be interpreted as representing the official policies, either expressed or implied, of the Army Research Office or the U.S. Government. The U.S. Government is authorized to reproduce and distribute reprints for Government purposes notwithstanding any copyright notation herein.

**Authors' contributions**

**Pulkit Garg**: Conceptualization, Methodology, Investigation, Writing – original draft, Writing – review & editing. **Daniel S. Gianola**: Conceptualization, Writing – review & editing, Funding acquisition. **Timothy J. Rupert**: Conceptualization, Writing – review & editing, Supervision, Project administration, Funding acquisition.

**Acknowledgments**

Not Applicable.

# References

I. Adlakha, P. Garg, and K. N. Solanki, Revealing the atomistic nature of dislocation-precipitate interactions in Al-Cu alloys. Journal of Alloys and Compounds **797**, 325 (2019)

M. R. Ahmadi, B. Sonderegger, E. Povoden-Karadeniz, A. Falahati, and E. Kozeschnik, Precipitate strengthening of non-spherical precipitates extended in ⟨100⟩ or {100} direction in fcc crystals. Materials Science and Engineering: A **590**, 262 (2014)

C. Antion, P. Donnadieu, F. Perrard, A. Deschamps, C. Tassin, and A. Pisch, Hardening precipitation in a Mg–4Y–3RE alloy. Acta Materialia **51**, 5335 (2003)

A. Ardell, V. Munjal, and D. Chellman, Precipitation hardening of Ni-Al alloys containing large volume fractions of γ′. Metallurgical Transactions A **7**, 1263 (1976)

A. J. Ardell, Precipitation hardening. Metallurgical Transactions A **16**, 2131 (1985)

D. Bacon, U. Kocks, and R. Scattergood, The effect of dislocation self-interaction on the Orowan stress. Philosophical Magazine **28**, 1241 (1973)

K. L. Baker and W. A. Curtin, Multiscale diffusion method for simulations of long-time defect evolution with application to dislocation climb. Journal of the Mechanics and Physics of Solids **92**, 297 (2016)

C. M. Barr, S. M. Foiles, M. Alkayyali, Y. Mahmood, P. M. Price, D. P. Adams, B. L. Boyce, F. Abdeljawad, and K. Hattar, The role of grain boundary character in solute segregation and thermal stability of nanocrystalline Pt–Au. Nanoscale **13**, 3552 (2021)

L. Brown and R. Ham, Dislocation-particle interactions. Strengthening methods in crystals, 9 (1971)

S. Chen, Z. H. Aitken, S. Pattamatta, Z. Wu, Z. G. Yu, D. J. Srolovitz, P. K. Liaw, and Y.-W. Zhang, Short-range ordering alters the dislocation nucleation and propagation in refractory high-entropy alloys. Materials Today **65**, 14 (2023)

Y. Cheng, E. Ma, and H. Sheng, Atomic level structure in multicomponent bulk metallic glass. Physical review letters **102**, 245501 (2009)

B. Clark, I. M. Robertson, L. Dougherty, D. Ahn, and P. Sofronis, High-temperature dislocation-precipitate interactions in Al alloys: An in situ transmission electron microscopy deformation study. Journal of materials research **20**, 1792 (2005)

D. Faken and H. Jónsson, Systematic analysis of local atomic structure combined with 3D computer graphics. Computational Materials Science **2**, 279 (1994)

Z. Feng, Y. Yang, B. Huang, M. Han, X. Luo, and J. Ru, Precipitation process along dislocations in Al–Cu–Mg alloy during artificial aging. Materials Science and Engineering: A **528**, 706 (2010)

D. Fougerouse, S. M. Reddy, M. Aylmore, L. Yang, P. Guagliardo, D. W. Saxey, W. D. A. Rickard, and N. Timms, A new kind of invisible gold in pyrite hosted in deformation-related dislocations. Geology **49**, 1225 (2021)

V. Gerold, On the structures of Guinier-Preston zones in AlCu alloys introductory paper. Scripta Metallurgica **22**, 927 (1988)

V. Gerold and H. Haberkorn, On the critical resolved shear stress of solid solutions containing coherent precipitates. physica status solidi (b) **16**, 675 (1966)

T. Gladman, Precipitation hardening in metals. Materials science and technology **15**, 30 (1999)

Y. N. Gornostyrev and M. Katsnelson, Misfit stabilized embedded nanoparticles in metallic alloys. Physical Chemistry Chemical Physics **17**, 27249 (2015)

P. Hirsch, The interpretation of the slip pattern in terms of dislocation movements. J. Inst. Metals **86**, 1957 (1957)

E. Hornbogen, Hundred years of precipitation hardening. Journal of Light Metals **1**, 127 (2001)

Y. Hu and T. J. Rupert, Atomistic modeling of interfacial segregation and structural transitions in ternary alloys. Journal of Materials Science **54**, 3975 (2019)

N. G. Kioussis and N. M. Ghoniem, Modeling of dislocation interaction with solutes, nano-precipitates and interfaces: A multiscale challenge. Journal of Computational and Theoretical Nanoscience **7**, 1317 (2010)

C. Kohler, P. Kizler, and S. Schmauder, Atomistic simulation of precipitation hardening in α-iron: influence of precipitate shape and chemical composition. Modelling and Simulation in Materials Science and Engineering **13**, 35 (2004)

V. S. Krasnikov, A. E. Mayer, and V. V. Pogorelko, Prediction of the shear strength of aluminum with θ phase inclusions based on precipitate statistics, dislocation and molecular dynamics. International Journal of Plasticity **128**, 102672 (2020)

M. E. Krug and D. C. Dunand, Modeling the creep threshold stress due to climb of a dislocation in the stress field of a misfitting precipitate. Acta Materialia **59**, 5125 (2011)

D. Kuhlmann-Wilsdorf, Theory of plastic deformation:-properties of low energy dislocation structures. Materials Science and Engineering: A **113**, 1 (1989)

M. Kuzmina, M. Herbig, D. Ponge, S. Sandlöbes, and D. Raabe, Linear complexions: Confined chemical and structural states at dislocations. Science **349**, 1080 (2015)

A. Kwiatkowski da Silva, G. Leyson, M. Kuzmina, D. Ponge, M. Herbig, S. Sandlöbes, B. Gault, J. Neugebauer, and D. Raabe, Confined chemical and structural states at dislocations in Fe-9wt%Mn steels: A correlative TEM-atom probe study combined with multiscale modelling. Acta Materialia **124**, 305 (2017)

C. Liu, S. K. Malladi, Q. Xu, J. Chen, F. D. Tichelaar, X. Zhuge, and H. W. Zandbergen, In-situ STEM imaging of growth and phase change of individual CuAlX precipitates in Al alloy. Scientific reports **7**, 2184 (2017)

H. Liu, B. Bellón, and J. Llorca, Multiscale modelling of the morphology and spatial distribution of θ′ precipitates in Al-Cu alloys. Acta Materialia **132**, 611 (2017)

I. Medouni, A. Portavoce, P. Maugis, P. Eyméoud, M. Yescas, and K. Hoummada, Role of dislocation elastic field on impurity segregation in Fe-based alloys. Scientific Reports **11**, 1780 (2021)

M. K. Miller, Atom probe tomography characterization of solute segregation to dislocations. Microscopy research and technique **69**, 359 (2006)

S. Mishra, S. Maiti, and B. Rai, Computational property predictions of Ta–Nb–Hf–Zr high-entropy alloys. Scientific Reports **11**, 4815 (2021)

J.-F. Nie, Precipitation and hardening in magnesium alloys. Metallurgical and Materials Transactions A **43**, 3891 (2012)

J. F. Nie and B. C. Muddle, Microstructural design of high-strength aluminum alloys. Journal of phase equilibria **19**, 543 (1998)

G. R. Odette, N. Almirall, P. B. Wells, and T. Yamamoto, Precipitation in reactor pressure vessel steels under ion and neutron irradiation: On the role of segregated network dislocations. Acta Materialia **212**, 116922 (2021)

E. Orowan, Symposium on internal stresses in metals and alloys. Institute of Metals, London **451** (1948)

S. P. Ringer and K. Hono, Microstructural Evolution and Age Hardening in Aluminium Alloys: Atom Probe Field-Ion Microscopy and Transmission Electron Microscopy Studies. Materials Characterization **44**, 101 (2000)

J. D. Robson, M. J. Jones, and P. B. Prangnell, Extension of the N-model to predict competing homogeneous and heterogeneous precipitation in Al-Sc alloys. Acta Materialia **51**, 1453 (2003)

B. Sadigh, P. Erhart, A. Stukowski, A. Caro, E. Martinez, and L. Zepeda-Ruiz, Scalable parallel Monte Carlo algorithm for atomistic simulations of precipitation in alloys. Physical Review B **85**, 184203 (2012)

C. V. Singh and D. H. Warner, Mechanisms of Guinier–Preston zone hardening in the athermal limit. Acta Materialia **58**, 5797 (2010)

C. V. Singh and D. H. Warner, An atomistic-based hierarchical multiscale examination of age hardening in an Al-Cu alloy. Metallurgical and Materials Transactions A **44**, 2625 (2013)

C. V. Singh, A. J. Mateos, and D. H. Warner, Atomistic simulations of dislocation–precipitate interactions emphasize importance of cross-slip. Scr. Mater. **64**, 398 (2011)

D. Singh, V. Turlo, D. S. Gianola, and T. J. Rupert, Linear complexions directly modify dislocation motion in face-centered cubic alloys. Materials Science and Engineering: A **870**, 144875 (2023)

W. Soboyejo, *Mechanical properties of engineered materials* (CRC press, 2002), Vol. 152.

M. Soleymani, M. H. Parsa, and H. Mirzadeh, Molecular dynamics simulation of stress field around edge dislocations in Aluminum. Computational Materials Science **84**, 83 (2014)

M. Starink and A. -M ZAHRA, Mechanisms of combined GP zone and θ′ precipitation in an Al-Cu alloy. Journal of materials science letters **16**, 1613 (1997)

A. Y. Stroev, O. Gorbatov, Y. N. Gornostyrev, and P. Korzhavyi, Solid solution decomposition and Guinier-Preston zone formation in Al-Cu alloys: A kinetic theory with anisotropic interactions. Physical Review Materials **2**, 033603 (2018)

A. Stukowski, Visualization and analysis of atomistic simulation data with OVITO–the Open Visualization Tool. Modelling and Simulation in Materials Science and Engineering **18**, 015012 (2009)

A. Stukowski, V. V. Bulatov, and A. Arsenlis, Automated identification and indexing of dislocations in crystal interfaces. Modelling and Simulation in Materials Science and Engineering **20**, 085007 (2012)

A. Takahashi and N. M. Ghoniem, A computational method for dislocation–precipitate interaction. Journal of the Mechanics and Physics of Solids **56**, 1534 (2008)

A. P. Thompson, H. M. Aktulga, R. Berger, D. S. Bolintineanu, W. M. Brown, P. S. Crozier, P. J. in 't Veld, A. Kohlmeyer, S. G. Moore, T. D. Nguyen, R. Shan, M. J. Stevens, J. Tranchida, C. Trott, and S. J. Plimpton, LAMMPS - a flexible simulation tool for particle-based materials modeling at the atomic, meso, and continuum scales. Computer Physics Communications **271**, 108171 (2022)

R. Thomson and R. Balluffi, Kinetic theory of dislocation climb. I. General models for edge and screw dislocations. Journal of Applied Physics **33**, 803 (1962)

V. Turlo and T. J. Rupert, Prediction of a wide variety of linear complexions in face centered cubic alloys. Acta Materialia **185**, 129 (2020)

A. Varma R., P. Pant, and M. P. Gururajan, Dislocation assisted phase separation: A phase field study. Acta Materialia **244**, 118529 (2023)

A. Vevecka-Priftaj, A. Böhner, J. May, H. W. Höppel, and M. Göken, in *Materials Science Forum* (Trans Tech Publ, 2008), pp. 741

C. Wolverton, First-principles prediction of equilibrium precipitate shapes in Al-Cu alloys. Philosophical magazine letters **79**, 683 (1999)

Y. Xiang and D. Srolovitz, Dislocation climb effects on particle bypass mechanisms. Philosophical magazine **86**, 3937 (2006)

S. Xu, D. L. McDowell, and I. J. Beyerlein, Sequential obstacle interactions with dislocations in a planar array. Acta Materialia **174**, 160 (2019)

S. Yin, Y. Zuo, A. Abu-Odeh, H. Zheng, X.-G. Li, J. Ding, S. P. Ong, M. Asta, and R. O. Ritchie, Atomistic simulations of dislocation mobility in refractory high-entropy alloys and the effect of chemical short-range order. Nature Communications **12**, 4873 (2021)

Z. Zhang, Z. Su, B. Zhang, Q. Yu, J. Ding, T. Shi, C. Lu, R. O. Ritchie, and E. Ma, Effect of local chemical order on the irradiation-induced defect evolution in CrCoNi medium-entropy alloy. Proceedings of the National Academy of Sciences **120**, e2218673120 (2023)

X. Zhou, J. R. Mianroodi, A. Kwiatkowski da Silva, T. Koenig, G. B. Thompson, P. Shanthraj, D. Ponge, B. Gault, B. Svendsen, and D. Raabe, The hidden structure dependence of the chemical life of dislocations. Science Advances **7**, eabf0563 (2021)